\documentclass[prl,twocolumn,showpacs]{revtex4}
\usepackage[dvips]{graphicx}

\newcommand{\ignore}[1]{}
\newcommand{\beq}{\begin{equation}}
\newcommand{\eeq}{\end{equation}}

\begin{document}

\title{Threshold Phenomena under Photo Excitation of Spin-crossover
Materials with Cooperativity due to Elastic Interactions}
\author{Seiji Miyashita$^{1,5}$, Per Arne Rikvold$^{2}$, 
Takashi Mori$^{1,5}$,
Yusuk\'e Konishi$^{3,5}$, 
Masamichi Nishino$^{4,5}$, 
Hiroko Tokoro$^{1,6}$
}
\affiliation{
$^{1}${\it Department of Physics, Graduate School of Science,
The University of Tokyo,7-3-1 Hongo, Bunkyo-Ku, Tokyo 113-8656, Japan}  \\
$^{2}$ {\it Department of Physics and Center for Materials Research and 
Technology, Florida State University, Tallahassee, Florida 32306-4350, USA} \\
$^{3}$ {Solid State Theory Division, Institute of Materials Structure 
Science, KEK, 1-1 Oho, Tsukuba, Ibaraki 305-0801, Japan} \\
$^{4}${\it Computational Materials Science Center, National Institute
for Materials Science, Tsukuba, Ibaraki 305-0047, Japan} \\
$^{5}${\it CREST, JST, 4-1-8 Honcho Kawaguchi, Saitama 332-0012, Japan}\\
$^{6}${\it PRESTO, JST, 4-1-8 Honcho Kawaguchi, Saitama 332-0012, Japan}
}
\date{\today}
\begin{abstract}
Photo-induced switching 
from the low-spin state to the high-spin state is 
studied in a model of spin-crossover materials, 
in which long-range interactions are induced by 
elastic distortions due to different molecular sizes the two spin states.
At a threshold value of the light intensity we observe nonequilibrium
critical behavior corresponding to a mean-field spinodal point.
Finite-size scaling of the divergence of the relaxation time is revealed 
by analysis of kinetic Monte Carlo simulations.   
\end{abstract}

\pacs{75.30.Wx 75.50.Xx 75.60.-d 64.60.-i}

\keywords{Mean field spinodal phenomena, Spin-crossover material,
photo-irradiation induced phase transition}
\maketitle


Molecular solids whose molecules can exist in two electro-vibrational states,
a low-spin ground state (LS) and a high-spin excited state (HS),
are known as spin-crossover (SC) compounds. Due to higher
degeneracy of the HS state, such materials
can be brought into a majority HS state at low temperatures 
by several methods, including light exposure,
known as Light-Induced Excited Spin-State Trapping (LIESST) \cite{GUTL94}. 
If the intermolecular interactions are sufficiently strong, this change
of state can become a discontinuous phase transition
such that the HS phase becomes metastable and hysteresis occurs.
The different magnetic
and optical properties of the two phases make such cooperative
SC materials promising candidates for switches, displays, and
recording media \cite{KAHN98}, and  
the cooperative properties of SC materials have 
therefore been studied extensively \cite{GUTL94}.
In addition to SC materials, models with similar long-range interactions are 
relevant to a range of physical phenomena, including earthquake faults 
\cite{KLEI07} and phase transitions in polymers \cite{BIND84}. 

In a typical model of a SC material, the energy difference between the 
HS and LS states is 
taken as $D >0$, while the degeneracy $g_{\rm HS}$ of HS 
is larger than $g_{\rm LS}$,
such that HS is preferable at high temperature. We can 
express the HS (LS) spin state of the $i$-th molecule by $\sigma_i=+1$ $(-1)$, 
giving the intra-molecule Hamiltonian,
\begin{equation}
{\cal H}_{\rm eff} = \frac{1}{2}\sum_{i} (D - k_{\rm B} T \ln g ) \sigma_i,
\label{WP0}
\end{equation}
where $g=g_{\rm HS}/g_{\rm LS}$.
Here the difference of the degeneracies are expressed by 
a temperature dependent field.
In order to provide the cooperative property in the SC transition, 
a short-range Ising-type interaction has previously been 
adopted \cite{Wajnflasz2,miyaPTP}. 
Recently, we pointed out an alternative mechanism with 
elastic interactions due to the 
volume difference between the LS and HS molecules 
\cite{Nishino2007,Konishi2007}.
The elastic interaction mediates the effects of local distortions over long
distances, and it has been found that the critical 
behavior belongs to the mean-field universality class \cite{miyaPRB2008}.

Here, we study the dynamics under photo-excitation from the perfect LS state
at low temperature. For a single molecule, we find a
smooth increase of the HS fraction, and the saturated HS fraction is a smooth
function of the irradiation intensity.
On the other hand, if we include a cooperative interaction,
a kind of threshold phenomenon appears.
When the irradiation is weak, the system stays in the LS state.
If the irradiation becomes stronger than a threshold value,
the system jumps to the HS state.
Thus, the stationary state suddenly changes from LS to HS at a threshold value 
of the irradiation intensity.
This sudden change is analogous to the sudden change of magnetization at 
the coercive field in the hysteresis loop of 
a ferromagnet following reversal of the applied field, 
known as a ``spinodal point."

The dynamics of field-driven first-order phase transitions 
have previously been studied in detail for Ising/lattice-gas 
models with short-range interactions. 
In that case, the nucleation and growth of clusters play important roles,
and different regimes have been identified, depending on the relative 
time scales of nucleation and growth \cite{RTMS}.
If the average time between nucleation events is long compared with the 
time it takes a single cluster to grow to a size comparable with the system, 
the transformation will happen stochastically in a Poisson 
process via a single cluster (single-droplet, or SD, regime), 
and the 
average lifetime of the metastable phase is inversely proportional to the 
system volume. 
If the average time between nucleation events is short compared with the 
growth time, many clusters nucleate while the first one is growing, resulting 
in an almost deterministic process 
(multi-droplet, or MD, regime), 
and the metastable lifetime is independent of the system volume 
(Avrami's law) \cite{RTMS}.

In this paper, we study the photo-induced transition in a model
of SC materials with elastic interactions, which we have developed 
previously \cite{Konishi2007,miyaPRB2008}.
In this model, we found that the appearance of large, compact clusters
is strongly suppressed, and the system remains  uniform
near the critical point and also during simulations of field-driven 
hysteresis \cite{miyaPRB2008}.
In the process of the photo-induced transition, we also expect to see 
effects of the effective long-range elastic interactions.
Here, we therefore investigate the relaxation
time near the threshold point and its dependence on the system parameters. 
We analyze the dependences from
the viewpoint of the mean-field spinodal, which
shows qualitatively different properties from field-driven phase 
transformation in short-range models \cite{UNGE84,GORM94}.


Here we use Monte Carlo (MC) simulations according to 
the constant-pressure method \cite{Konishi2007}.
In addition, we include a process of photo-excitation by the
following procedure.
We choose a site randomly and change it to the HS state with a
probability $a$ if it is previously in the LS state.   
We denote $a$ as the irradiation intensity. 

The order parameter for the present model is the 
fraction of HS molecules,
$f_{\rm HS}$,  which is expressed by the ``magnetization" 
\beq
M = {1\over N}\sum_i^Ns_i = \frac{1}{2}(f_{\rm HS} -1).
\eeq 
Here we adopt the two-dimensional model of Ref.~\cite{miyaPRB2008} with 
the same parameters, including  
$g=20$ and $D=0.5$. 

\begin{figure}[t]
\centerline{\includegraphics[clip,width=5cm]{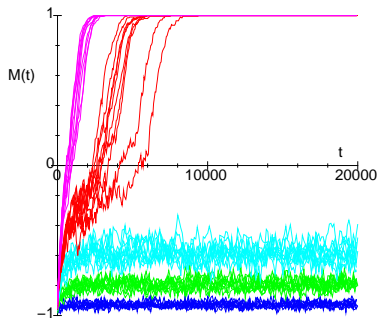} }
\caption{(Color online) 
Time series of the order parameter $M(t)$ under irradiation. 
Irradiation intensities $a=0.001$, 0.002, 0.003, 0.004,
and 0.005 from bottom to top. We plot 10 samples for
each value of $a$. The system size is $L=50$.
The time $t$ is measured in Monte Carlo steps per site (MCSS). 
}
\label{Fig_Mtime}
\end{figure}

\begin{figure}[t]
\centerline{\includegraphics[clip,width=4.0cm]{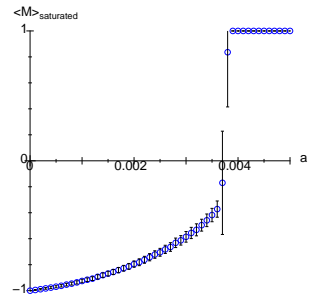} }
\caption{(Color online) 
Saturated order parameter $M$, proportional to the HF fraction, 
shown vs the irradiation strength $a$.
}
\label{Fig_saturate}
\end{figure}

In Fig.~\ref{Fig_Mtime}, we depict typical time series of the 
HS fraction under irradiation.
We find a sudden change in the saturated values. 
For $a\le 0.003$ the system stays in the LS state, 
while it jumps to the HS state 
for $a\ge 0.004$. In Fig.~\ref{Fig_saturate}, we show 
the dependence of the saturated value on $a$.
Because of the cooperative interaction, the HS fraction remains
low when the irradiation is sufficiently weak.
On the other hand, it approaches unity
when $a$ exceeds a threshold value.
This threshold value of $a$ is estimated from Fig.~\ref{Fig_saturate} 
to be slightly below 0.004.

\begin{figure}[t]
$$\begin{array}{cc}
\includegraphics[clip,width=4cm]{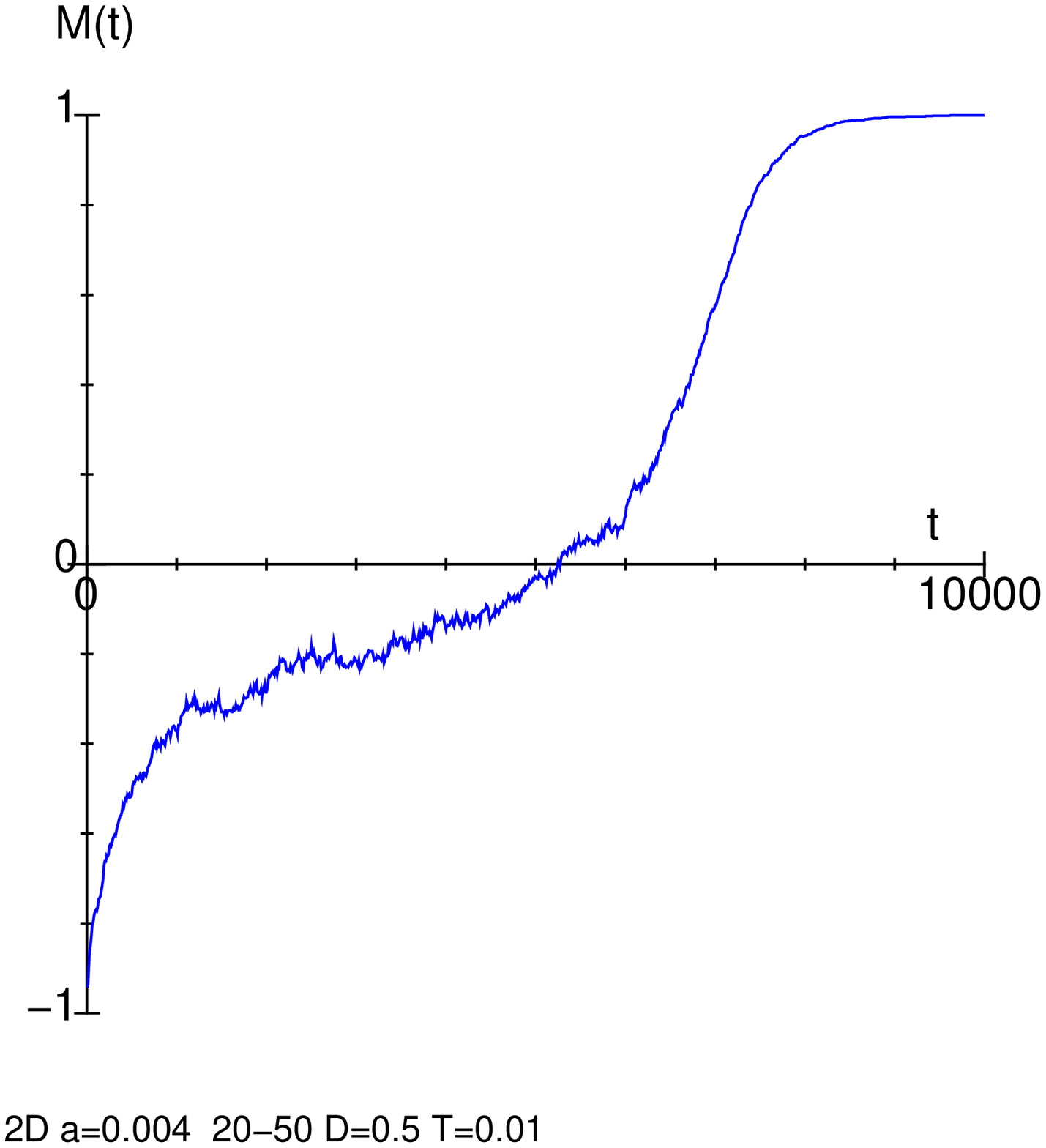} &
\includegraphics[clip,width=3cm]{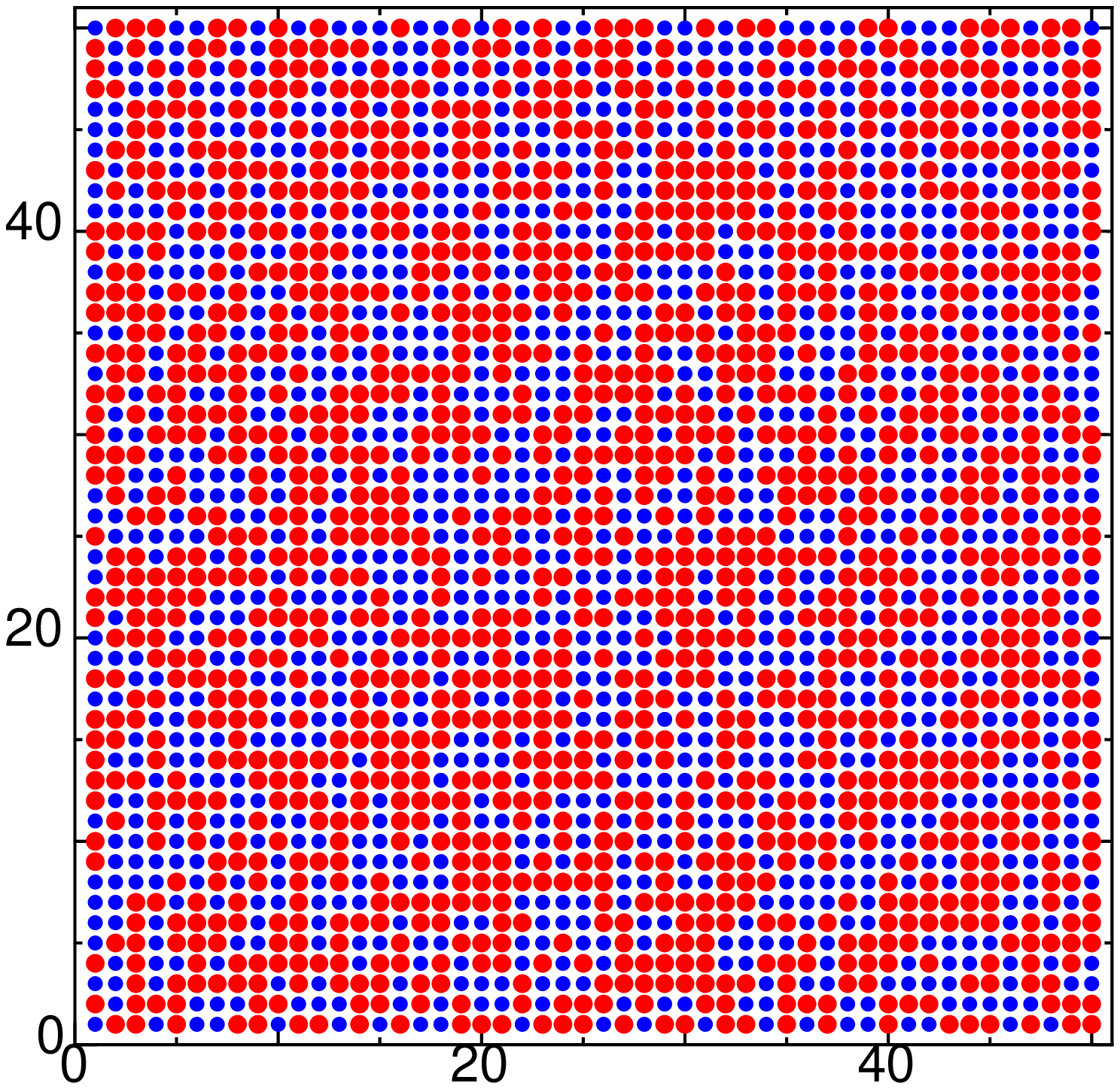}\\
({\rm a}) & ({\rm b})\end{array} $$ 
\caption{(Color online) 
(a)
An example of the time evolution of the order parameter $M(t)$ under 
photo irradiation with $a=0.004$. 
Time is given in MCSS. 
(b) Spin configuration during the process of photo irradiation
shown in (a) at $t=6000$ MCSS.
}
\label{single}
\end{figure}

We also find a sudden change of $M$ as a function of time.
We depict an example of the relaxation process for $L=50$ in 
Fig.~\ref{single}(a),
which shows a sharp change around $t=5000$ Monte Carlo steps per site (MCSS).
The HS fraction stays at an intermediate value ($M\simeq -0.2$) 
for a while, and then it rapidly increases to the saturated value. 
This sudden change is reminiscent of a transformation via single-droplet
nucleation from the metastable phase to the equilibrium phase.
However, no visible clusters are observed. 
In Fig.~\ref{single}(b) we depict a configuration at $t=6000$ MCSS, near the  
starting point of the sudden change. This absence of cluster structures is the 
same characteristic which we found previously in 
simulations of hysteresis in this model \cite{miyaPRB2008}.

\begin{figure}[t]
$$\begin{array}{c}
\includegraphics[clip,width=5.0cm]{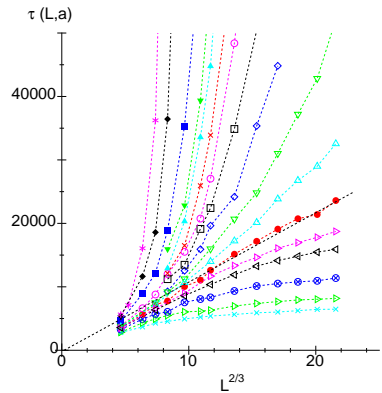} \\
\end{array} 
$$
\caption{(Color online) 
Dependence of the relaxation time $\tau$ on system size $L$ 
and irradiation strength $a$. 
Shown as $\tau$ vs $L^{2/3}$ for $a=$ 
0.00365, 0.00370, 0.00375,
0.00377, 0.00378, 0.00379,
0.00380, 0.00381, 0.00382,
0.00383, 0.00384, 0.00385, 
0.00386, 0.00397, 0.00390,
0.00395 and 0.00400 
from above to below. 
Straight-line behavior consistent with
(\protect\ref{N667}) is seen for $a= 0.00385 \approx a_{\rm SP}$.
}
\label{RXN667}
\end{figure}

Next, we study the system-size dependence of the relaxation time.
Here we define the relaxation time $\tau$ as the time 
at which the average of $\sigma_i$
reaches 0.5, corresponding to a HS fraction of 0.75, 
starting from the perfect LS spin state ($\sigma_i=-1$).
We perform $N_{\rm sample} (\ge 200)$ independent runs 
for each parameter set and average the results.
In Fig.~\ref{RXN667}, we depict the size dependences of the relaxation
times $\tau$ for $a \in  [0.00365, 0.00400]$, obtained by averaging over the
samples. Here, we plot the data as a function of $L^{2/3}$ because it is 
expected that the relaxation time diverges as $L^{2/3}$ at the spinodal 
point (see below). 

We found the sudden change of the saturated value of $M$ around 
$a_{\rm SP} \approx 0.0385$ in Fig.~\ref{Fig_saturate}.
Below this value, the relaxation times increase rapidly with $L$, while
above it the relaxation times saturate for large $L$.
As shown in Fig.~\ref{Fig_Mtime}, for $a < a_{\rm SP}$ 
the times of the jump are widely distributed.
This is a sign of stochastic nucleation, although we 
do not observe any clusters in Fig.~\ref{single}(b).
For $a > a_{\rm SP}$, the transformation process becomes nearly deterministic.

It is also found that the size dependence of the 
standard deviation of the relaxation time 
$\sqrt{\langle\tau^2\rangle-\langle\tau\rangle^2}$ 
is very different below and above the threshold.
Below $a_{\rm SP}$ the standard deviation is about the same as the 
mean value while it decreases with increasing $L$ above $a_{\rm SP}$.
The former suggests a kind of Poisson process in the SD region and 
the latter a kind of Gaussian process in the MD region.
In Table I, we list typical examples.

\begin{table}
\caption{List of average and the standard deviation of the distribution of
the relaxation time.
}
\begin{tabular}{rrrrr}\hline
$a$ & $L$ & $N_{\rm sample}$ & $\langle\tau\rangle$ & 
$\sqrt{\langle\tau^2\rangle-\langle\tau\rangle^2}$ \\ \hline
 0.00375 &  20 &   200 &       12174.045 &        8189.569 \\
 0.00375 &  40 &   200 &      105239.460 &       96223.716 \\
 0.00375 &  60 &   400 &     4608237.905 &     4343268.886 \\
 0.00400 &  20 &   200 &        4281.680 &        1321.136 \\
 0.00400 &  40 &   200 &        5394.570 &        1111.389 \\
 0.00400 &  60 &   200 &        5905.740 &         945.039 \\
 0.00400 &  80 &   600 &        6245.863 &         727.554 \\
 0.00400 & 100 &   350 &        6475.271 &         591.310 \\
\hline
\end{tabular}
\end{table}

It should be noted that the relaxation time increases with the system size.
This system-size dependence of the relaxation time is {\it qualitatively\/} 
different from the behavior in systems with short-range 
interactions \cite{RTMS}.
In the latter case, the relaxation time is determined by the probability 
of creating a critical {\it localized, compact\/} droplet,
whose free energy is {\it independent\/} of the system size. 
Droplets thus nucleate in a Poisson process of rate $L^d I$, where 
$d$ is the spatial dimensionality of the system (here, $d=2$). 
Thus, the relaxation time becomes short when the size increases
in both the SD and MD regions.

Compared to this dependence in short-range systems, the increase of the 
relaxation time in the present system with the 
system size is notable. The divergence when $L\rightarrow\infty$
indicates that the present threshold phenomenon is not just 
a crossover as in short-range systems, but a true
critical phenomenon. Thus we will study its critical properties by analyzing 
the size dependence of the relaxation behavior. 

The simple van Hove type dynamics using
the mean-field approximation implies that the waiting time before  
a jump over the free-energy barrier for $a < a_{\rm SP}$ has the form 
\beq
\tau\sim \exp(cN),
\label{ecN}
\eeq
where $N=L^d$ is the volume of the system and $c$ is positive.
The present case appears similar. 
Indeed, the present threshold behavior is essentially the same as the 
sudden phase change near a mean-field spinodal point, 
and we expect that the parameter dependence 
is the same as that near the spinodal point of the  
Ising ferromagnet with weak long-range interactions 
(the Husimi-Temperley model), with the irradiation strength 
$a$ here playing the role of the magnetic field.
Thus, the relaxation time for $a > a_{\rm SP}$ should diverge as \cite{BIND73} 
\beq
\tau\propto (a-a_{\rm SP})^{-1/2},
\label{diverge}
\eeq
independent of $L$. 
At the spinodal point, the relaxation time diverges with the system size as 
\beq
\tau\propto N^{1/3},
\label{N667}
\eeq
where $N$ is the total number of spins \cite{KMK,SP,SPmori}.

\begin{figure}[t]
\centerline{\includegraphics[clip,width=4.5cm]{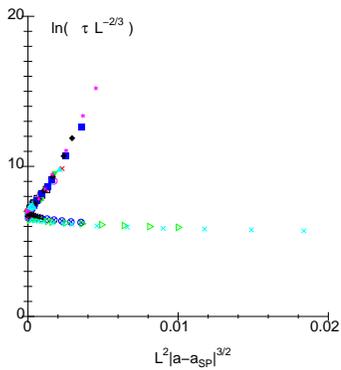}}
\caption{(Color online) 
Scaling plot of the scaling function for the relaxation time, $\ln f(x)$ with 
$x^{3/2}$. 
Here we adopt $a_{\rm SP}=0.00385$. 
The symbols denote the data of the same values of $a$ as in Fig.~\ref{RXN667}.
}
\label{tauscaling}
\end{figure}

We plot the relaxation times for various values of $a$ 
as functions of $N^{1/3}=L^{2/3}$ in Fig.~\ref{RXN667}.
The relaxation times 
at $a=0.00385$ are seen to follow a straight line, consistent
with the properties of the mean-field spinodal point (\ref{N667}).
The relations (\ref{ecN}), (\ref{diverge}), and (\ref{N667}) 
can be combined into a scaling form,
\beq
\tau\propto L^{d/3}f(L^{2d/3}(a-a_{\rm SP})),
\label{scaling1}
\eeq
where the scaling function $f(x)$ has the following asymptotic dependences 
on the scaling variable $x = L^{4/3}(a-a_{\rm SP})$:
\beq
f(x) \sim 
\left\{
\begin{array}{ll}
\exp\left(|x|^{3/2} \right) & \mbox{for $x \ll -1$}\\
x^0                        & \mbox{for $|x| \ll 1$}\\
x^{-1/2}                   & \mbox{for $x \gg 1$}
\end{array}
\right.
.
\label{scaling2}
\eeq
In Fig.~\ref{tauscaling}, we plot the scaling function $f(x) = \tau/L^{2/3}$ 
on logarithmic scale vs $|x|^{3/2}$ in order to have all the data including 
the large relaxation times at small values of $a$ in one figure.
We find that all the points collapse onto a scaling form, which shows the
asymptotic behavior given in (\ref{scaling2}).
This scaling form has been found for the spinodal phenomenon 
in the Husimi-Temperley model as well \cite{SPmori}. 

\begin{figure}[t]
\centerline{\includegraphics[clip,width=4.5cm]{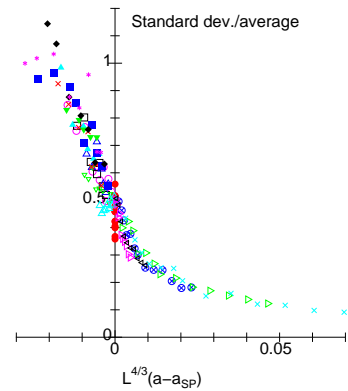}}
\caption{(Color online) 
Scaling plot of the ratio of the standard 
deviation and the mean of the relaxation time with 
$x = L^{4/3}(a-a_{\rm SP})$ with $a_{\rm SP}=0.00385$.
The symbols denote the data of the same values of $a$ as in Fig.~\ref{RXN667}.
}
\label{fig:ratio}
\end{figure}

We also study the finite-size scaling of the standard deviation 
of the distribution of the relaxation time.
Above $a_{\rm SP}$, the distribution of the relaxation time is
narrowly localized, and the standard deviation decreases with 
increasing system size.
On the other hand, far below $a_{\rm SP}$, the distribution  
is exponential, so that the standard deviation is 
of the same order as the average. 
In Fig.~\ref{fig:ratio} we show the ratio of the standard deviation to the 
mean of the relaxation time vs the scaling variable $x$. The results 
collapse quite well onto a scaling function. 

In conclusion we note that the present spin-crossover model with elastic 
interactions displays a sharp spinodal point with critical properties 
consistent with those of the mean-field universality class. 
We hope that this truly critical nonequilibrium phenomenon soon will be 
experimentally observed during photo-irradiation of SC materials.  

This work was partially supported by a Grant-in-Aid for Scientific 
Research on Priority Areas
``Physics of new quantum phases in superclean materials" 
(Grant No.\ 17071011), 
Grant-in-Aid for Scientific Research (C) (20550133) from MEXT, 
and also by the Next Generation Super Computer Project, 
Nanoscience Program of MEXT.
Numerical calculations were done on the supercomputer of ISSP. 
P.A.R.\ gratefully acknowledges hospitality at The University of Tokyo.
Work at Florida State University was supported by U.S.\ 
National Science Foundation Grants No.\ DMR-0444051 and DMR-0802288.

\end{document}